\documentclass[conference]{IEEEtran}
\IEEEoverridecommandlockouts
\usepackage{cite}
\usepackage{amsmath,amssymb,amsfonts}
\usepackage{algorithmic}
\usepackage{graphicx}
\usepackage{textcomp}
\usepackage{xcolor}
\usepackage{times}
\usepackage{helvet}
\usepackage{courier}
\usepackage{xurl}
\usepackage{hyperref}
\usepackage{enumitem}
\setlist{nosep}

\def\BibTeX{{\rm B\kern-.05em{\sc i\kern-.025em b}\kern-.08em
    T\kern-.1667em\lower.7ex\hbox{E}\kern-.125emX}}
\begin{document}

\title{Explainable Lightweight Compact Deep Models for Speech Emotion Recognition\\
}

\author{\IEEEauthorblockN{Nelly Elsayed}
\IEEEauthorblockA{\textit{School of Information Technology} \\
\textit{University of Cincinnati}\\
Ohio, United States\\
elsayeny@ucmail.uc.edu
}
}

\maketitle

\begin{abstract}
Speech Emotion Recognition (SER) is an important component in a wide range of human-centered applications, including healthcare, customer service, and human–computer interaction. In medical and decision-support settings, there is increasing interest in models that not only achieve accurate emotion recognition but also support transparent predictions and efficient deployment. However, many existing SER approaches rely on complex deep learning architectures that limit interpretability and increase computational cost.
This paper presents an explainable and lightweight speech emotion recognition framework based on a compact convolutional neural network architecture. The proposed approach utilizes log-Mel spectrogram representations to capture spectro-temporal speech characteristics and employs attentive statistics pooling to emphasize emotionally salient temporal segments. To improve model transparency, gradient-based class activation mapping (Grad-CAM) is incorporated to visualize the time–frequency regions that influence the model’s predictions.
Experimental evaluation on the SAVEE emotional speech dataset demonstrates that the proposed framework achieves competitive recognition performance while maintaining a compact architecture with significantly fewer parameters than many existing SER models. The results indicate that efficient convolutional architectures combined with interpretable analysis can provide a practical balance between recognition accuracy, computational efficiency, and model transparency.
\end{abstract}


\section{Introduction}

\begin{figure}
    \centering
    \includegraphics[width=1.0\linewidth]{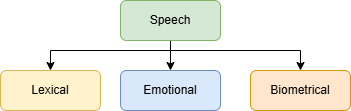}
    \caption{Speech signal components based on the information that each component holds.}
    \label{speech_contents}
\end{figure}

Speech is one of the primary ways people communicate, exchange thoughts, discuss their ideas, and interact with one another. Speech is more than just linguistic content. As shown in Figure~\ref{speech_contents}, speech combines lexical content, which is described by the spoken words and their contextual meaning, emotional content, which reflects the emotions expressed during speech, and biometric content which is characterized by vocal features that distinguish one speaker from another.

Speech emotion recognition (SER) has been rapidly advancing due to several factors, including the rapidly growing medical interest in a better understanding of mental health disorders for diagnostics and monitoring purposes~\cite{jordan2025speech,elsayed2022speech}. From an application design perspective, incorporating emotional factors into human-computer interaction can enhance overall user satisfaction when using a particular application. From a business perspective, SER can enhance user experience through personalized applications and communication that fulfill client requests while considering their emotional state~\cite{dar2024speech}.

There are two major ways to address the SER problem based on the speech data representation: image-based, and time–frequency approaches~\cite{latif2021survey}. The image-based SER approach aims to investigate the SER as an image classification task. First raw speech signals are transformed two-dimensional (2D) time–frequency representations, such as Mel spectrograms, log-Mel spectrograms, Mel-Frequency Cepstral Coefficient (MFCC) heatmaps, or spectral contrast maps~\cite{pawar2021convolution,elsayed2025empirical}. These representations are then classified using deep learning models designed for image analysis. This approach is widely used because it enables the application of powerful deep learning architectures that have demonstrated strong learning capabilities and high classification accuracy~\cite{chen2024vesper,hassan2024benchmarking,zhang2021pre}. In addition, image-based representations can capture emotion-discriminative spectro-temporal patterns. However, on the other side, this approach is computationally expensive due to the required feature extraction step and data transformation. Thus, this approach is not preferred for real-time applications or deployment on small computing devices such as mobile platforms, Internet of Things (IoT), and Internet of Medical Things (IoMT) devices.

The time–frequency SER approach addresses the SER problem using waveforms or low-level descriptors that are processed directly as one-dimensional time-series data. This approach typically uses one-dimensional convolutional neural networks (1D CNNs), temporal architectures such as long short-term memory (LSTM) or gated recurrent unit (GRU) networks, temporal convolutional networks (TCNs), and lightweight transformer models~\cite{ishaq2023tc,elsayed2025litelstm,etienne2018cnn,zhao2019speech,elsayed2022speech,abdelhamid2022robust}. This approach is significantly more efficient for edge and mobile devices because it involves substantially lower preprocessing overhead, making it computationally cheaper than image-based SER approaches. Moreover, it is conceptually closer to the natural representation of speech signals, as it operates directly on the original signal rather than a transformed image representation. However, this approach is more challenging to train compared to the image-based approach due to the lack of large-scale benchmarks and pretrained models for speech signals, which can increase training time and make it more difficult to achieve high recognition accuracy and reliable classification performance.

Despite recent progress, many SER studies emphasize predictive performance using deep or hybrid architectures, with less attention to deployment efficiency, reproducibility under speaker-independent evaluation, and qualitative inspection of the acoustic regions influencing model predictions. As a result, there remains a need for SER frameworks that balance recognition performance with computational efficiency while also enabling inspection of the spectro-temporal cues used by the model.

Rather than introducing a fundamentally new learning architecture, this work focuses on the practical integration of compact convolutional modeling and post-hoc visualization within a deployment-oriented SER pipeline.

To address these challenges, this paper presents a lightweight speech emotion recognition framework designed for resource-constrained environments. The proposed approach combines compact convolutional modeling with post-hoc visualization techniques to analyze the spectro-temporal regions that contribute to emotion predictions.

The main contributions of this paper are summarized as follows:

\begin{itemize}
\item We present a compact CNN-based speech emotion recognition pipeline using log-Mel spectrogram representations designed for low-resource inference environments.
\item We examine the trade-off between recognition performance and model complexity using a lightweight convolutional architecture.
\item We incorporate Grad-CAM-based post-hoc visualization to analyze the spectro-temporal regions associated with model predictions.
\end{itemize}

\section{Related Work}

\subsection{Deep Learning for Speech Emotion Recognition}
Deep learning models have been widely adopted for speech emotion recognition due to their ability to capture complex acoustic patterns. CNN-based models effectively learn spectro-temporal structures, while recurrent architectures such as LSTM or GRU networks capture temporal dependencies in speech signals~\cite{zhang2018speech,zhao2019speech}. However, hybrid architectures often introduce substantial computational overhead, limiting their suitability for resource-constrained environments~\cite{schuller2018speech}.

\subsection{Lightweight SER for Edge and Mobile Devices}
As many SER applications are deployed on edge or mobile devices, several studies have explored lightweight SER models that target limited computational resources in real time. Compact CNN architectures with reduced depth and parameter counts have been proposed to balance performance and efficiency. For example, depthwise separable convolutions and global pooling layers have been employed to significantly reduce computational complexity while maintaining competitive recognition accuracy~\cite{deng2020introducing}. Other studies explore lightweight recurrent architectures such as GRU networks, or compact convolutional architectures such as 1D CNNs, to accelerate processing time at lower cost than the LSTM~\cite{elsayed2022speech}. Other approaches prioritize simplified architectures and compact feature representations, such as low-dimensional Mel-spectrograms or short temporal contexts. While these models demonstrate feasibility for on-device emotion recognition, many focus primarily on efficiency–accuracy trade-offs while providing limited insight into the internal decision-making behavior of the models~\cite{chen2018speech}. 
These limitations motivate the development of SER frameworks that balance computational efficiency with interpretable model behavior.

\subsection{Explainability in SER}

Explainability in SER remains relatively underexplored compared to other speech processing tasks~\cite{jayasinghe2025systematic}. Only a limited number of studies have applied post-hoc explanation techniques, such as saliency maps or attention visualization, to analyze emotion-related speech cues. Gradient-based methods, including Grad-CAM, have recently been adopted to visualize discriminative spectro-temporal regions in CNN-based SER models~\cite{giacomelli2025large,selvaraju2017grad}.

However, explainability is often treated as an auxiliary analysis rather than an integrated component of the modeling pipeline. In addition, existing explainable SER approaches frequently rely on complex architectures, which limits their suitability for deployment on edge or mobile devices~\cite{elsayed2026socio}. These limitations highlight the need for lightweight SER frameworks that enable inspection of model behavior while maintaining computational efficiency.

Building on these observations, this work investigates whether a compact convolutional architecture combined with post-hoc visualization techniques can provide an efficient and interpretable SER framework suitable for deployment in resource-constrained environments.

\section{Methodology}

\subsection{Feature Extraction}
For the SER task, compact time--frequency representations are extracted as log-Mel spectrograms, which are widely used in SER due to their ability to preserve emotion-relevant spectral and temporal information while remaining computationally efficient~\cite{lu2025fusion}. 

Given a discrete-time speech signal $x[n]$ sampled at frequency $f_s$, the signal is segmented into overlapping frames using window length $W$ and hop size $H$. The short-time Fourier transform (STFT) is then computed as:
\begin{equation}
    X(n, k) = \sum_{m=0}^{W-1} x[m + nH] \, w[m] \, e^{-j \frac{2\pi}{W} k m},
\end{equation}
\noindent
where $w[m]$ is the windowing function, $n$ is the frame index, $m$ is the sample index within a frame, and $k$ is the frequency-bin index.

The magnitude-squared STFT is then projected onto a Mel-scaled filterbank consisting of $F$ triangular filters:
\begin{equation}
M(n, f) = \sum_{k} \left| X(n, k) \right|^2 \, \Phi_f(k),
\end{equation}
\noindent
where $f \in \{1, 2, \ldots, F\}$ indexes the Mel filters, and $\Phi_f(k)$ denotes the response of the $f^{\mathrm{th}}$ Mel filter at frequency bin $k$.

Logarithmic compression is applied to reduce the dynamic range of the representation:
\begin{equation}
S(n, f) = \log \left( M(n, f) + \epsilon \right),
\end{equation}
\noindent
where $\epsilon$ is a small positive constant added for numerical stability.

All utterances are padded or cropped to a fixed temporal duration, resulting in a spectrogram $S \in \mathbb{R}^{F \times T}$, where $T$ denotes the number of time frames. Per-utterance normalization is then applied to reduce speaker-dependent and recording-dependent variability. The resulting normalized log-Mel spectrogram $\hat{S}$ is used as input to the proposed compact SER model.

\subsection{Compact Convolutional Neural Network Backbone}

The proposed SER framework employs a lightweight convolutional backbone designed for efficient deployment on resource-constrained devices. In contrast to deep hybrid architectures that combine multiple heavy modules, the proposed model uses a shallow convolutional design to balance representational capacity and computational cost.

The normalized log-Mel spectrogram is first passed through a convolutional stem layer, followed by a sequence of compact convolutional blocks. Each block applies convolution, batch normalization, nonlinear activation, and dropout to progressively learn discriminative spectro-temporal patterns associated with emotional expression. To reduce model size and computation, the architecture uses compact convolutional processing with a small number of feature channels and limited depth. This design preserves the two-dimensional time--frequency structure of the input while maintaining low memory and inference overhead.

The final convolutional feature maps encode local emotion-relevant patterns across both time and frequency dimensions. Rather than directly flattening these maps or using large fully connected layers, the model transforms them into a compact sequence of frame-level representations that can be efficiently aggregated for utterance-level classification.

\subsection{Attentive Statistics Pooling}

To aggregate frame-level representations into a fixed-length utterance-level embedding, the proposed model employs attentive statistics pooling (ASP). Unlike global average pooling, which treats all temporal frames equally, ASP learns attention weights that emphasize emotionally informative temporal regions.

Let $\mathbf{h}_t \in \mathbb{R}^{C}$ denote the frame-level embedding at time step $t$, where $t = 1,2,\ldots,T'$. An attention mechanism assigns a normalized importance weight $\alpha_t$ to each frame such that:
\begin{equation}
    \sum_{t=1}^{T'} \alpha_t = 1.
\end{equation}

Using these learned weights, ASP computes an attention-weighted mean vector:
\begin{equation}
    \boldsymbol{\mu} = \sum_{t=1}^{T'} \alpha_t \mathbf{h}_t,
\end{equation}
and an attention-weighted standard deviation vector:
\begin{equation}
    \boldsymbol{\sigma} = \sqrt{\sum_{t=1}^{T'} \alpha_t (\mathbf{h}_t - \boldsymbol{\mu})^2 }.
\end{equation}

The final utterance-level representation is then formed by concatenating the first- and second-order statistics:
\begin{equation}
    \mathbf{z} = [\boldsymbol{\mu}; \boldsymbol{\sigma}].
\end{equation}

In the proposed framework, the convolutional feature maps are first averaged along the frequency axis to produce a sequence of temporal feature vectors. These vectors are then projected into a lower-dimensional embedding space and aggregated using ASP. This mechanism allows the model to focus on emotionally salient segments while reducing the influence of silence, padded regions, or less informative frames. In addition, the learned attention weights provide a natural temporal interpretability signal that can be visualized during analysis.

\subsection{Explainability via Gradient-Based Activation Mapping}

Explainability is incorporated into the proposed SER framework through gradient-based class activation mapping (Grad-CAM)~\cite{selvaraju2017grad}. Grad-CAM is used as a post-hoc visualization technique to identify the time--frequency regions that contribute most strongly to a predicted emotion class.

Specifically, given a target class, Grad-CAM computes the gradients of the class score with respect to the feature maps of the final convolutional layer. These gradients are globally averaged to obtain channel-wise importance weights, which are then combined with the corresponding convolutional feature maps to produce a class-specific activation map. The resulting heatmap highlights the regions of the learned representation that are most influential for the prediction.

In this work, the Grad-CAM heatmap is upsampled and overlaid on the input log-Mel spectrogram to support qualitative inspection of spectro-temporal patterns associated with the model’s decision. This explainability component is intended to aid interpretation of model behavior and does not directly modify the classification objective or improve predictive performance.

\begin{figure}
    \centering
    \includegraphics[width=1.0\linewidth]{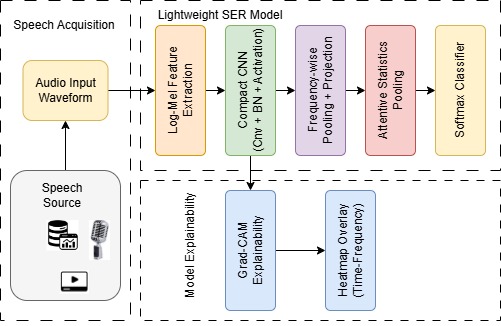}
    \caption{The proposed explainable and lightweight SER framework.}
    \label{fig:framework}
\end{figure}

\subsection{Proposed Framework}

The proposed lightweight and explainable SER framework is illustrated in Figure~\ref{fig:framework}. The framework integrates compact feature extraction, efficient convolutional modeling, attentive temporal aggregation, and post-hoc explainability within a unified pipeline intended for resource-constrained environments.

As shown in Figure~\ref{fig:framework}, raw speech signals are first converted into normalized log-Mel spectrograms that preserve the joint temporal and spectral structure of emotional speech while maintaining a compact input representation. These spectrograms are then processed by a lightweight convolutional backbone that learns local spectro-temporal patterns relevant to emotion recognition.

The output feature maps of the final convolutional block are reduced along the frequency dimension to form a sequence of frame-level feature vectors. These vectors are projected and aggregated through attentive statistics pooling, which produces an utterance-level representation by emphasizing informative temporal segments and modeling their statistical variability. The resulting representation is passed to the final classifier to generate the predicted emotion label.

In parallel, an explainability module based on Grad-CAM is applied to the final convolutional layer to visualize the spectro-temporal regions that contribute most strongly to each prediction. These class-specific activation maps are overlaid on the corresponding log-Mel spectrograms to support qualitative analysis of the model’s attention to emotion-relevant acoustic structure.

Overall, the proposed framework is designed to jointly support efficient inference and interpretable analysis. By combining compact convolutional modeling with attentive temporal pooling and Grad-CAM-based visualization, the framework provides a practical SER pipeline suitable for edge, mobile, and wearable deployment scenarios.

\section{Experimental Setup}

\subsection{Dataset}

To evaluate the proposed framework, experiments are conducted on the Surrey Audio-Visual Expressed Emotion (SAVEE) dataset~\cite{vlasenko2007combining}. The dataset contains recordings from four male speakers expressing seven emotions: anger, disgust, fear, happiness, sadness, surprise, and neutrality. In total, the dataset includes 480 utterances, with 120 sentences recorded per speaker.
Each utterance consists of a phonetically balanced sentence originally selected from the TIMIT corpus~\cite{haq2008audio}. The audio recordings are sampled at 16 kHz and contain expressive emotional speech captured in a controlled recording environment.
To reduce the risk of speaker leakage and ensure fair evaluation, a speaker-independent protocol is used. Specifically, the dataset is partitioned according to speaker identity, where recordings from different speakers are assigned to the training, validation, and test sets. This protocol ensures that speakers appearing in the test set are not observed during training.
In the experiments, three speakers are used for training and validation, while the remaining speaker is used for testing. This configuration ensures that the evaluation is performed on previously unseen speakers.

\subsection{Model Setup}

The proposed SER model operates on fixed-length log-Mel spectrograms extracted from speech recordings sampled at 16 kHz. The short-time Fourier transform (STFT) is computed using an FFT size of 400 samples, corresponding to a window length of 25 ms, and a hop size of 10 ms (160 samples). The resulting spectra are projected onto 64 Mel filters to produce log-Mel spectrogram representations.
To maintain consistent input dimensions, all utterances are padded or truncated to a maximum duration of four seconds. This results in a fixed-size time–frequency representation used as input to the neural network. The classification model consists of a compact convolutional neural network backbone followed by attentive statistics pooling (ASP). The convolutional backbone contains three lightweight convolutional blocks designed to capture local spectro-temporal patterns while maintaining a low parameter count.
The model parameters are optimized using the Adam optimizer with a learning rate of $2 \times 10^{-3}$ and a weight decay of $10^{-3}$. Training is performed using mini-batches of size 32 for up to 160 epochs. To ensure reproducibility, all experiments are conducted using a fixed random seed controlling data shuffling and model initialization.

\subsection{Results and Analysis}

Table~\ref{tab:model_summary} summarizes the training characteristics of the proposed model, including training time, average time per epoch, number of trainable parameters, and training accuracy. The compact architecture contains only 33,208 trainable parameters, which is substantially smaller than most deep learning SER models reported in the literature.

\begin{figure*}
    \centering
    \includegraphics[width=13cm, height = 3.5cm]{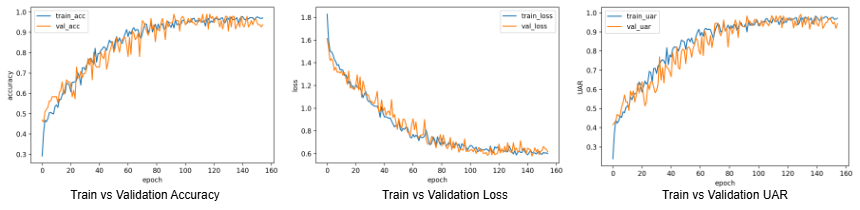}
    \caption{The proposed SER model train versus validation accuracy, loss, and UAR diagrams.}
    \label{fig:plots}
\end{figure*}

Figure~\ref{fig:plots} illustrates the training and validation curves for accuracy, loss, and unweighted average recall (UAR). The curves demonstrate stable convergence during training and show limited signs of overfitting, indicating that the lightweight architecture can effectively learn emotion-related speech patterns despite its compact size.

\begin{table}[t]
    \centering
     \caption{The proposed lightweight SER model summary.}
    \begin{tabular}{|l|l|}
    \hline
    \textbf{Metrics}& \textbf{Value}\\
    \hline
       Train time  & 245.87 sec.\\
  
        Avg. time/epoch & 2.08 sec.\\

        No. Trainable param. & 33208\\
        Train acc. & 98.33\%\\
        \hline
    \end{tabular}
   
    \label{tab:model_summary}
\end{table}

\begin{figure}[t]
    \centering
    \includegraphics[width=1.0\linewidth]{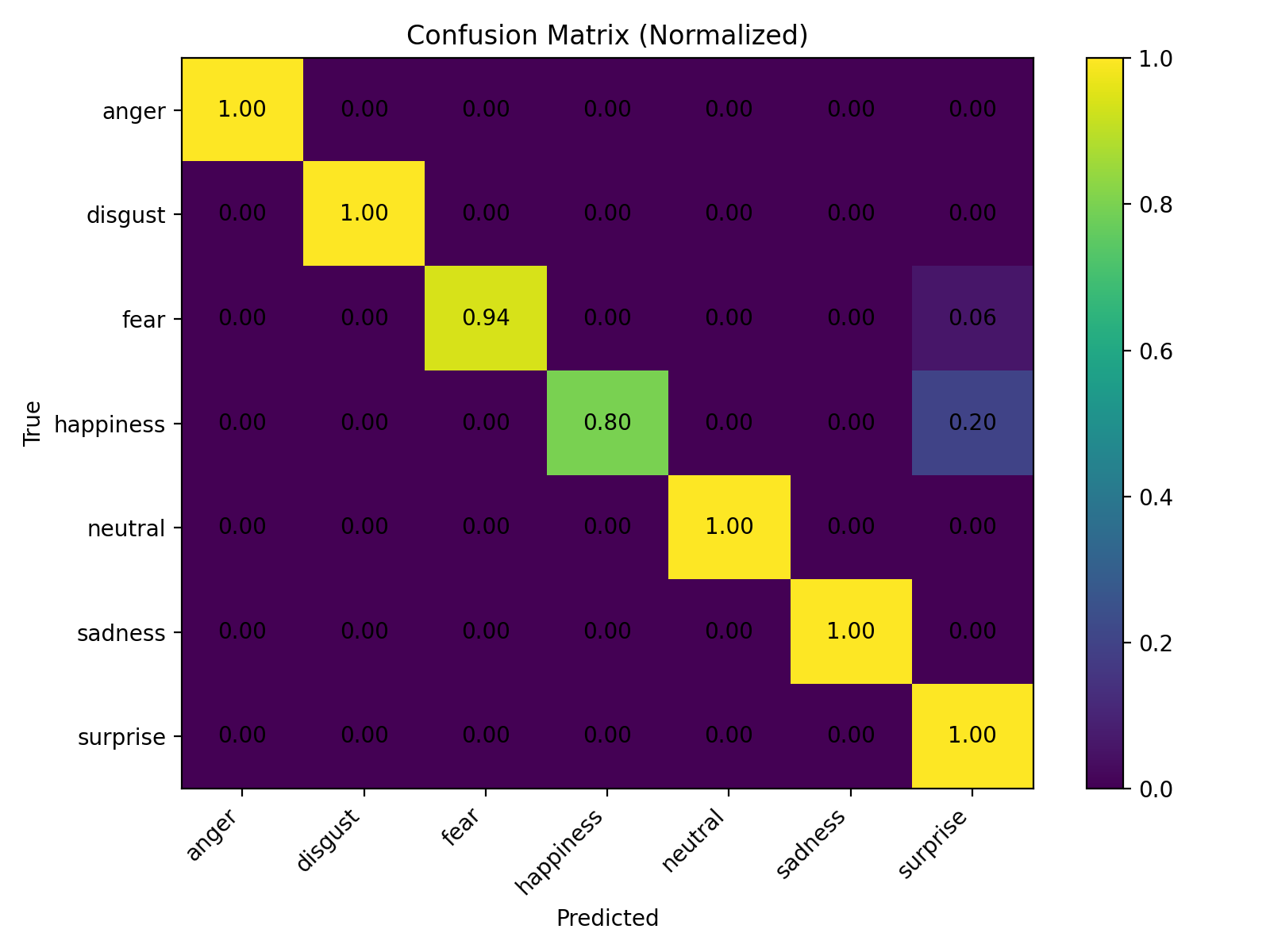}
    \caption{The proposed explainable and lightweight SER confusion matrix over SAVEE dataset.}
    \label{fig:confusion_normalized}
\end{figure}

The normalized confusion matrix shown in Figure~\ref{fig:confusion_normalized} provides insight into the model’s class-level behavior. Most emotional categories are correctly classified with high recall, indicating that the model effectively captures emotion-discriminative acoustic patterns. Some confusion occurs between emotions with similar acoustic characteristics, which is consistent with prior SER studies where emotions such as sadness and neutral speech share overlapping prosodic features.

\begin{table}[t]
	\caption{Testing performance of the proposed lightweight SER model on the SAVEE dataset.}
	\begin{center}
		\begin{tabular}{|l|l|}
			\hline
			\textbf{Metrics}& \textbf{Value} \\
			\hline
            Accuracy & 96.875\%\\
            UAR & 0.977\\
            Kappa & 0.96184\\
            95\% CI & (0.93, 1.00)\\
            Hamming Loss& 0.03125 \\
            F1-score& 0.96875\\
            FPR & 0.0052 \\
            FNR & 0.0313 \\
            TPR (Recall) & 0.9688 \\
            TNR (Specificity) & 0.9948 \\

			\hline
		\end{tabular}
		\label{fig:testing}
	\end{center}
\end{table}

The testing results summarized in Table~\ref{fig:testing} demonstrate that the proposed lightweight SER model achieves strong recognition performance across multiple evaluation metrics. In addition to overall accuracy, the high unweighted average recall (UAR) indicates balanced recognition performance across emotion classes, which is particularly important for emotion recognition tasks where class distributions may vary.

The reported 95\% confidence interval further indicates that the observed performance remains consistently high across the evaluation samples, suggesting stable generalization behavior of the model. The high Cohen’s Kappa score also confirms strong agreement between predicted and true emotion labels beyond chance-level classification.

\begin{table*}[h]
\caption{Comparison between the proposed lightweight SER model and representative deep learning-based approaches evaluated on the SAVEE dataset.}
\begin{center}
\begin{tabular}{|p{4.5cm}|p{3cm}|p{3.9cm}|p{1cm}|l|}
\hline
\textbf{Model} & \textbf{Feature Representation} & \textbf{Architecture} & \textbf{Acc. (\%)} & \textbf{\# Parameters} \\
\hline

Mountzouris et al.~\cite{mountzouris2023speech} & Log-Mel spectrogram & CNN + Attention (ATN) & 74.0 & $\sim$0.5--2 M \\

Liu et al.~\cite{Liu2024CNNA_LSTM_Fusion} & Spectrogram features & CNN + Attention + LSTM fusion & 94.5 & $\sim$3--15 M \\

Ali et al.~\cite{Ali2023HybridMFCCTCNN} & Hybrid MFCC features & CNN classifier & 93.0 & $\sim$0.2--2 M \\

Mishra et al.~\cite{Mishra2024HybridCNNBiLSTM} & eGeMAPS features & Deep Belief Network & 56.76 & $\sim$0.1--2 M \\

Pinnamaraju et al.~\cite{ResNetSAVEEChapter2024} & Spectrogram features & ResNet (transfer learning) & 85--90 & $\sim$11--26 M \\

Ottoni et al.~\cite{ottoni2023deep} & Spectrogram features & Meta-learned CNN & 90.62 & $\sim$0.8--3 M \\

Ahmed et al.~\cite{10.1007/978-3-031-47994-6_14} & Spectrogram features & 4-layer CNN & $\sim$76 & $\sim$1 M \\

Nath et al.~\cite{Nath2024SAVEE_ML_Comparative} & Handcrafted features & Random Forest (ML baseline) & 72 & $\sim$0.5 M \\

Nath et al.~\cite{Nath2024SAVEE_ML_Comparative} & Acoustic features & BiLSTM & 72 & $\sim$5 M \\

Ouyang et al.~\cite{ouyang2025speech} & Spectrogram features & CNN-LSTM hybrid & 61.1 & $\sim$1--5 M \\

\textbf{Proposed Lightweight SER (Ours)} & Log-Mel spectrogram & \textbf{Compact CNN + ASP + XAI} & \textbf{96--97} & \textbf{0.033 M (33,208)} \\
\hline

\end{tabular}
\label{tab:compare}
\end{center}

\footnotesize{Parameter counts marked with ranges are approximate estimates derived from model descriptions when exact values were not reported in the original publications.}
\end{table*}

Because prior studies often employ different experimental protocols, feature representations, and dataset splits, direct numerical comparisons should be interpreted with caution. Table~\ref{tab:compare} therefore provides a high-level comparison with representative SER models reported in the literature that have been evaluated on the SAVEE dataset.

As shown in Table~\ref{tab:compare}, the proposed lightweight SER model is compared with several state-of-the-art deep learning approaches. Since many of these studies do not explicitly report the number of trainable parameters, the parameter counts listed in the table are approximate estimates derived from the model descriptions provided in the corresponding publications. 

The results indicate that the proposed lightweight model achieves competitive recognition performance while maintaining a substantially lower parameter count compared with many existing SER models. These findings suggest that compact architectures can provide an effective balance between recognition performance and computational efficiency, making them suitable for deployment in resource-constrained environments.

\begin{figure}
    \centering
    \includegraphics[width=1.0\linewidth]{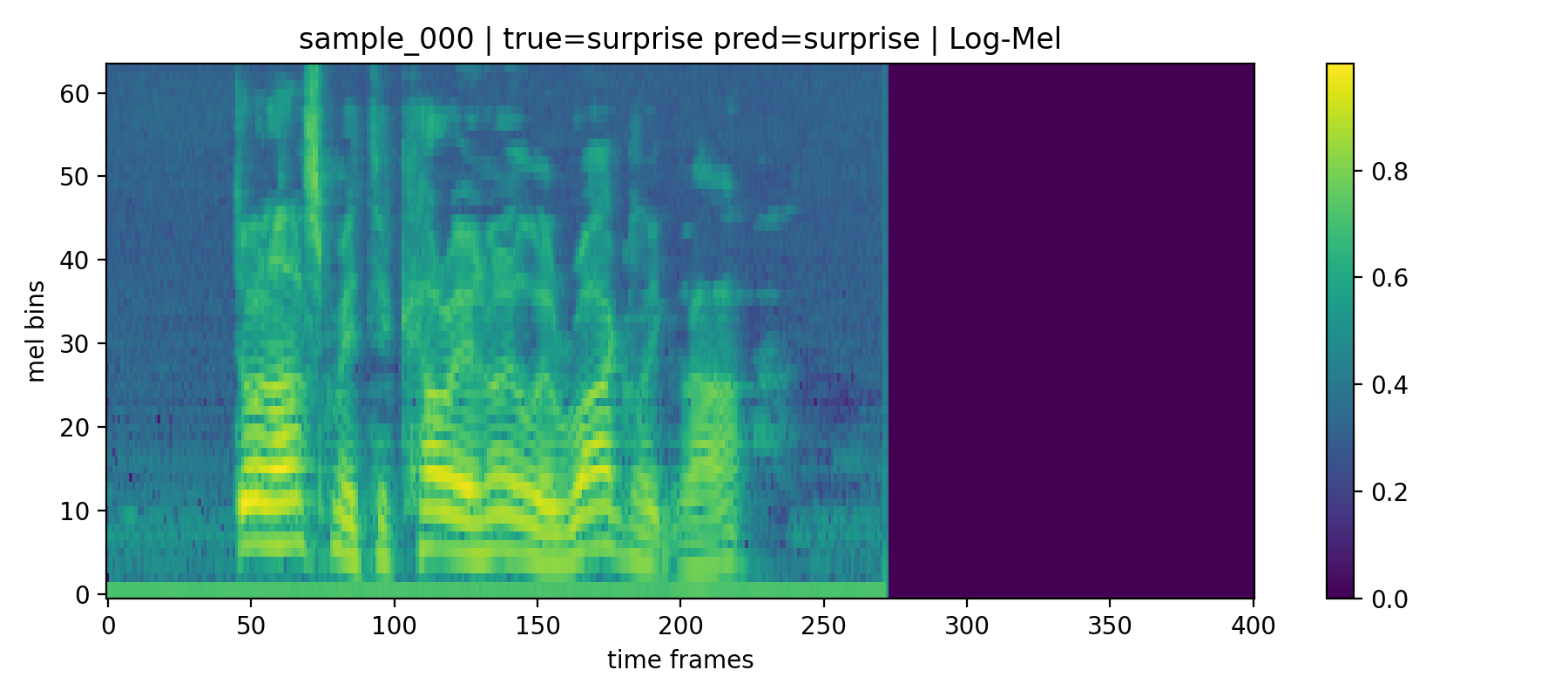}
    \caption{The Log-Mel spectrogram of a SAVEE utterance classified as surprise. The horizontal axis denotes time frames and the vertical axis denotes Mel frequency bins. Zero-padding is applied after the speech segment to obtain a fixed-length input.}
    \label{fig:log_mel}
\end{figure}

\begin{figure}[t]
    \centering
    \includegraphics[width=1.0\linewidth]{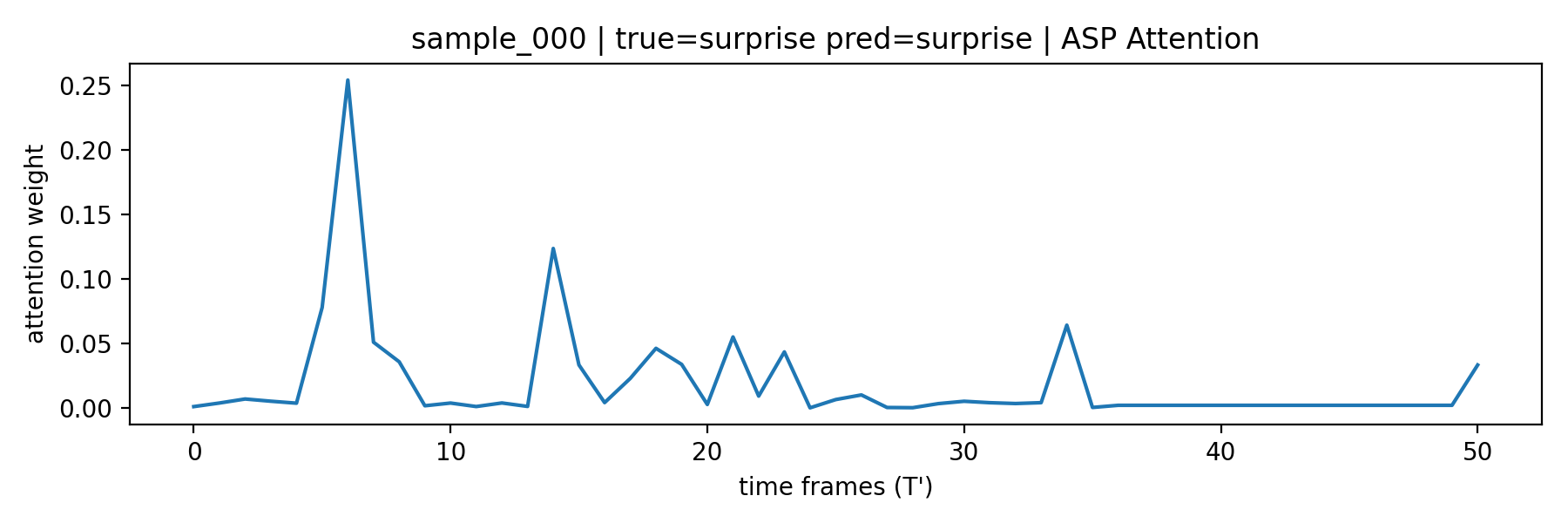}
    \caption{Temporal attention weights produced by the attentive statistics pooling (ASP) module. Each value corresponds to the relative contribution of a downsampled time frame to the final utterance-level representation.}
    \label{fig:attentive}%
\end{figure}

\begin{figure}[t]
    \centering
    \includegraphics[width=1.0\linewidth]{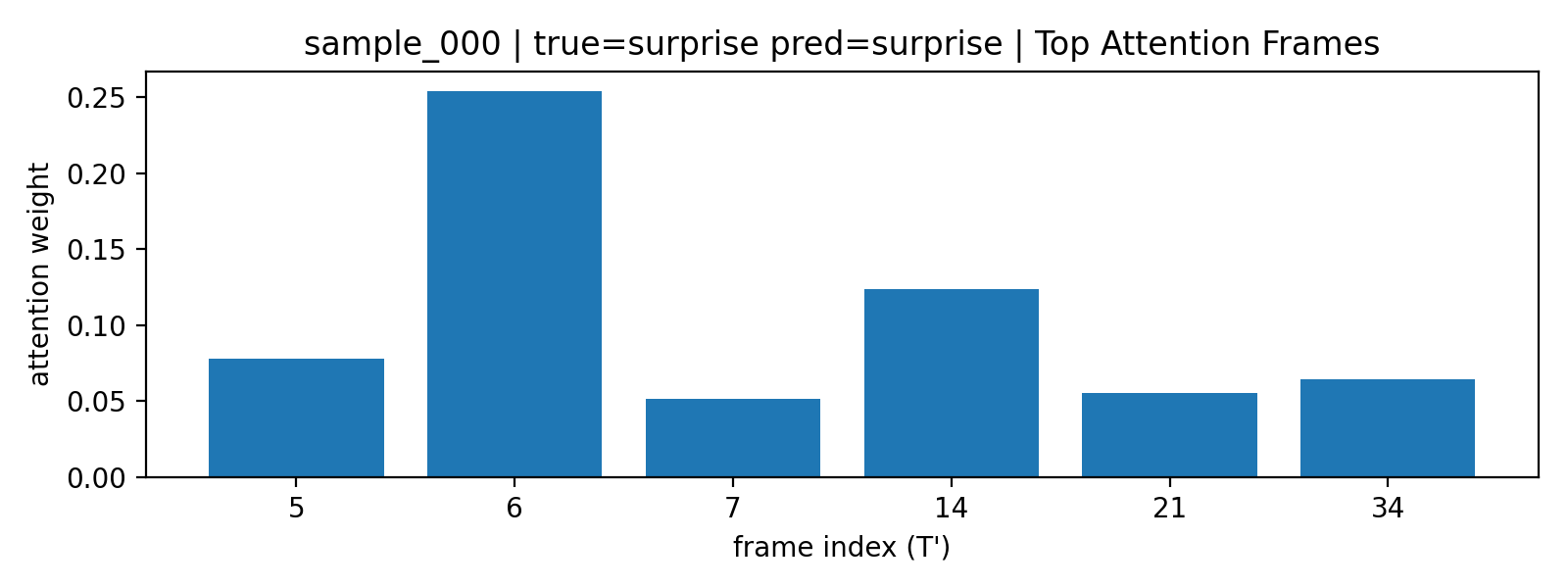}
    \caption{Top-K time frames selected according to the highest ASP attention weights. These frames represent the most influential temporal segments used by the model when forming the utterance-level embedding.}
    \label{fig:topk}%
\end{figure}

\begin{figure}
    \centering
    \includegraphics[width=1.0\linewidth]{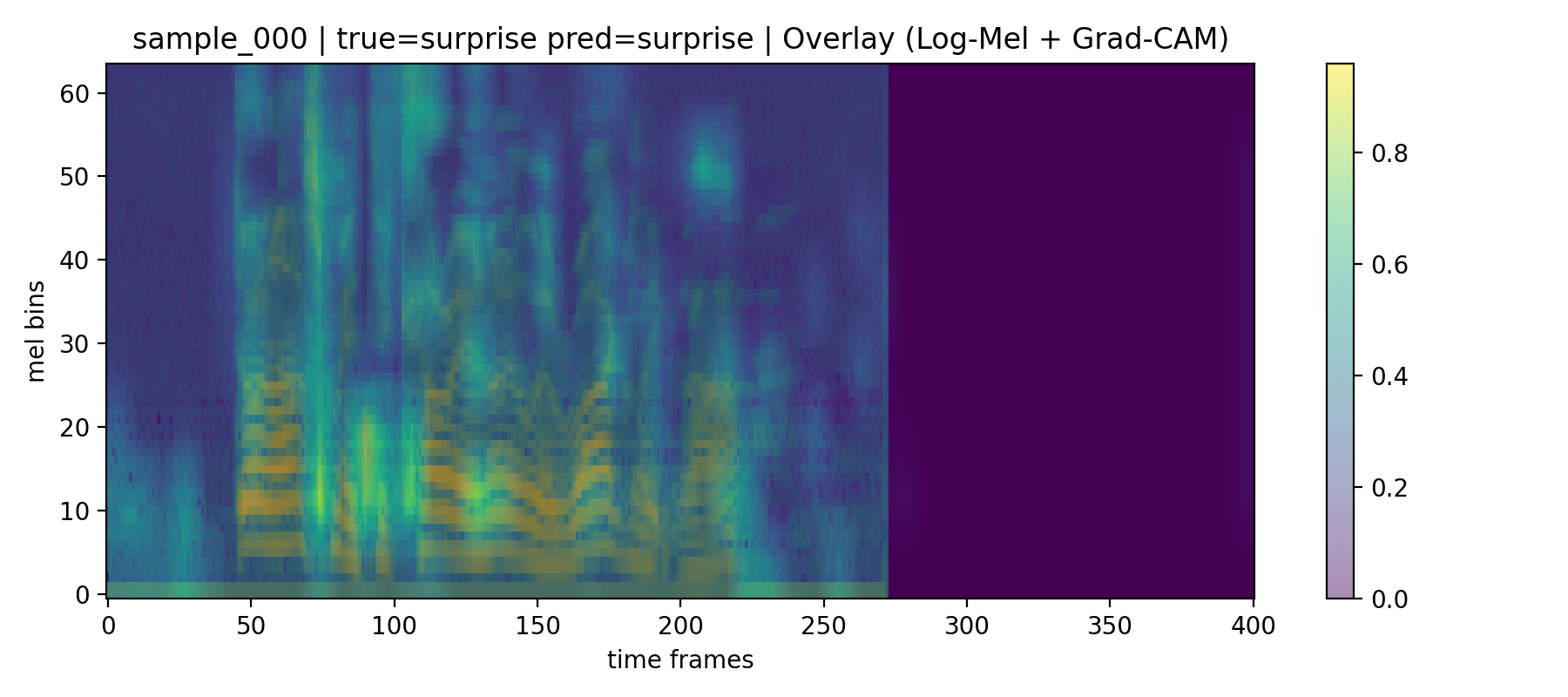}
    \caption{Overlay of Grad-CAM activations on the log-Mel spectrogram. The alignment between high-activation regions and acoustically salient speech components.}%
    \label{fig:overlay}
\end{figure}




\subsection{Explainability Analysis}

To support interpretability, the proposed framework incorporates multiple explainability mechanisms that provide insight into the model's decision-making behavior.

Figure~\ref{fig:log_mel} illustrates the log-Mel spectrogram of an example utterance labeled as \textit{surprise}. The rightmost dark region corresponds to zero-padding added to enforce a fixed input duration. The absence of activation in this padded region confirms that the model correctly ignores non-informative frames.

Temporal importance is analyzed using the attentive statistics pooling (ASP) mechanism. As shown in Figure~\ref{fig:attentive}, the learned attention weights highlight specific time frames that contribute most strongly to the final prediction. These peaks typically correspond to emotionally expressive segments characterized by changes in pitch, energy, or spectral structure.

To further summarize the temporal importance, the top-$k$ attention frames are extracted as illustrated in Figure~\ref{fig:topk}. These frames represent the most influential speech segments used by the model when forming the utterance-level embedding.

At the feature level, Grad-CAM is applied to visualize discriminative spectro-temporal regions learned by the convolutional layers. Figure~\ref{fig:overlay} shows that the strongest activations occur in low-to-mid frequency regions during expressive speech intervals. The alignment between Grad-CAM activations and acoustic speech structure indicates that the model focuses on meaningful emotional cues rather than spurious patterns.

These explainability analyses provide qualitative insight into the temporal and spectral cues used by the model while confirming that emotion predictions are driven by interpretable acoustic patterns.

\section{Conclusion}

This paper proposes an explainable, lightweight deep learning framework for speech emotion recognition, specifically designed for deployment on resource-constrained environments such as edge, mobile, and wearable devices. By leveraging compact convolutional neural network architectures operating on log-Mel spectrograms, the proposed approach achieves a strong balance between recognition performance, computational efficiency, and model transparency.

Experimental results on the SAVEE benchmark demonstrate that the proposed model attains high classification accuracy and unweighted average recall while requiring only 33,208 trainable parameters. Compared with existing deep learning-based SER approaches, the proposed framework achieves competitive or superior performance with significantly lower model complexity, making it well-suited for real-time and low-power applications.

This paper presented an explainable and lightweight speech emotion recognition framework designed for resource-constrained environments. The proposed approach combines compact convolutional modeling with attentive statistics pooling and Grad-CAM-based visualization to support both efficient inference and interpretable analysis. Experimental evaluation on the SAVEE dataset demonstrates that the model achieves strong recognition performance while maintaining a compact architecture with only 33k parameters. These results suggest that effective SER systems can be developed using lightweight architectures without sacrificing interpretability or deployment efficiency.

\bibliographystyle{ieeetr}
\bibliography{references_speech}
\end{document}